\documentclass[twoside,11pt]{article}
\usepackage{amsmath,xcolor,colortbl,comment}

\usepackage{subfig}
\usepackage{csquotes}
\usepackage{float}

\usepackage{jmlr2e}

\jmlrheading{1}{2021}{1-48}{18/09}{NaN}{Straccamore}{Matteo Straccamore, Luciano Pietronero, Andrea Zaccaria}
\ShortHeadings{Which will be your firm's next technology?}{Straccamore, Pietronero and Zaccaria}
\firstpageno{1}

\begin{document}

\title{Which will be your firm's next technology? \\\Large Comparison between machine learning and network-based algorithms}

\author{\name Matteo Straccamore \email matteo.straccamore@cref.it \\
       \addr Dipartimento di Fisica Universit\`a ``Sapienza"\\
       P.le A. Moro, 2, 00185 Rome, Italy\\
       Centro Ricerche Enrico Fermi\\
       Piazza del Viminale, 1, 00184 Rome, Italy
       \AND
       \name Luciano Pietronero \\
       \addr Centro Ricerche Enrico Fermi\\
       Piazza del Viminale, 1, 00184 Rome, Italy
       \AND
        \name Andrea Zaccaria \\
       \addr Istituto dei Sistemi Complessi (ISC) - CNR\\
       UoS Sapienza,P.le A. Moro, 2, 00185 Rome, Italy\\
       Centro Ricerche Enrico Fermi\\
       Piazza del Viminale, 1, 00184 Rome, Italy
       }

\editor{//}

\maketitle

\begin{abstract}
We reconstruct the innovation dynamics of about two hundred thousand companies by following their patenting activity for about ten years. We define the technological portfolios of these companies as the set of the technological sectors present in the patents they submit. By assuming that companies move more frequently towards related sectors, we leverage on their past activity to build network-based and machine learning algorithms to forecast the future submission of patents in new sectors. We compare different evaluation metrics and prediction methodologies, showing that tree-based machine learning algorithms overperform the standard methods based on networks of co-occurrences. This methodology can be applied by firms and policymakers to disentangle, given the present innovation activity, the feasible technological sectors from those that are out of reach, given their present innovation activity.
\end{abstract}

\keywords{Economic Complexity, Technological Innovation, Predictions, Patenting firms}

\flushbottom
\maketitle
\thispagestyle{empty}

\section*{Introduction}
The question regarding the nature of the link between the performance of firms and their internal allocation of resources \citep{penrose1959the} and capabilities \citep{teece1994understanding} has fueled the interest of economics and management scholars for a long time, since opening the black box of corporate strategy would be key to gain insight into the determinants of corporate heterogeneity and hence a better understanding of markets and their evolution. To the best of our knowledge, these analyses are all aiming at finding explanatory variables for the present performance and not at forecasting future activity. On the contrary, the approach known as Economic Fitness and Complexity, widely applied at both country and regional level, naturally focuses on forecasting, which represent a natural, scientifically sound framework to validate and falsify the different approaches \citep{tacchella2018dynamical,albora2021product,tacchella2021relatedness}. The aim of the present paper is to apply the EFC forecasting methods at firm level, and in particular to the bipartite network of firms and the technological sectors in which they show patenting activity. \\ 
One of the main problems for the economical literature is to empirically track the capabilities and the strategic choices of companies. Unfortunately, these elements are generally intangible, so that the empirical literature often struggles to find instruments to keep up with the theoretical richness of the debate.
One of the more easily measurable footprints left behind by the strategic decision making of firms is \textit{diversification}, i.e the scope of activities (both at technological and productive level) to which internal resources are devoted.
This has been recognized early by scholars, who have often focused their efforts in this direction to reconcile theory with empirical evidence \citep{penrose1960growth, gort1962diversif, berry1971diversif}.
Though, diversification is interesting in and of itself, perhaps the more interesting question regards the degree of complementarity (or relatedness) between the various elements included in the portfolio of activities in which businesses engage.
Notable early efforts to address this aspect have been proposed by \cite{rumelt1974strategy} and \cite{rumelt1982diversification}, whose focus was centered on the nexus between profitability and the degree of relatedness between the business units of the same corporation to test the hypothesis that higher profitability correlates, in diversified manufacturing firms, with expanding primarily into areas share a core skill or resource.\\
\cite{teece1994understanding} have built on the above intuition by employing plant-level data classifying establishments according to the standard 4-digit SIC industrial codes relative to the industrial sectors in which they operate and measuring the relatedness between sectors through the frequency of their co-occurrence within the same productive plant, that is two sectors are related if many plants produce both.
The hypothesis underlying this approach is the so-called \emph{survivor principle} \citep{teece1994understanding}, \emph{i.e.} the assumption that economic competition eventually drives inefficient organizational forms out of the market, thus promoting the co-occurrence of activities that are well integrated with one another because of complementarities in the \emph{technological capabilities} they require.
In virtue of the survivor principle, efficient combinations of activities should occur with a significantly higher frequency than one would expect if activities were paired randomly.
Indeed, the authors find that internal coherence matters, as firms that diversify tend to add activities that are related to at least a part of their existing portfolio. More recent analyses confirmed this hypothesis
\citep{rahmati2020all,buccellato2016competences,lo2016firms}.\\
Production is not the only aspect of corporate strategy in which building a coherent portfolios of related activities has been shown to matter (for example in \cite{gort1962diversification, rumelt1974strategy, berry1971corporate} the manufacturing sector is considered).
Indeed, in the last twenty years, the empirical analysis of the innovative output of firms as measured by patents has gained increasing popularity \citep{rycroft1999complexity}.
It is worth noting that patent data have become in general a workhorse for the literature on technical change over the past few decades due to the growing availability of machine-readable patent documents and widespread access to sufficient computing power \citep{youn2015invention}.
All the above has played a pivotal role in fueling this trend spurring scholarly (e.g. \cite{hall2001nber}),
institutional (e.g. PATSTAT, REGPAT) and corporate (e.g Google Patents) efforts aimed at constructing comprehensive collections of patent-related documents.
Increasing data availability has in turn allowed researchers to inquire into the nature of patented inventions, their role in explaining technical change, their reciprocal connections, and their link to inventor - and applicant-specific characteristics \citep{strumsky2011measuring,strumsky2012using,youn2015invention}.
One of the characteristics of patent documents, which historically has lent itself more to economic analysis, is the presence of codes associated with the claims contained in the patent applications. Claims are the claims that mark the boundary of the commercial exclusion rights demanded by inventors. To allow evaluation by patent office examiners, claims are classified based on the technological areas they impact according to classifications (e.g. the IPC classification \citep{fall2003automated}), which consist of a hierarchy of 6-digit codes that associate progressively finer-grained definitions of technological areas to the codes lower in the hierarchy.
Mapping claims to classification codes allows to localize patents and patent applications within the technology space.
Taking advantage of the increasing availability of patent data, several studies \citep{jaffe2000knowledge, leten2007technological, joo2010measuring, rigby2015technological} have found significant empirical evidence suggesting that evidence that relatedness in the composition of R\&D activities has implications for the ability of firms to innovate successfully.\\
Within this stream of literature, a well-known study \citep{breschi2003knowledge} has recovered the methodology proposed by \cite{teece1994understanding} and built upon it to investigate whether firms tend to diversify their innovative efforts in a coherent fashion by patenting in technological fields that share a common knowledge base with the technological fields in which they innovated in the past.
In particular, the authors have analyzed the technological diversification of firms through the co-occurrences between technology codes.\\
In another well-known paper, \cite{nesta2006firm} have studied corporate knowledge coherence in the US pharmaceutical industry showing that both the scope and the coherence of the knowledge base ``contribute positively and significantly to the firm's innovative performance'', as measured by the number of patents it produces weighted by the number of citations received.\\
Some authors of the present paper introduced the concept of ``coherent diversification'' \citep{pugliese2019coherent2}, showing that firms that diversify (i.e., expand their technological portfolios by patenting in a relatively large number of technology sectors) in a coherent way (i.e., by preferring related sectors to unrelated ones) on average show a higher performance in terms of labor productivity. Here, the relatedness between technology sectors is measured by suitably normalized co-occurrences.\\ 
Finally, we mention the work by \cite{kim2021technological}, who have studied the relatedness between technology codes in Korean firms, finding that "firms are more likely to develop a new technology when they already have related technologies".\\
In this work we quantify the relatedness between a firm and a technology sector in different ways, namely using both standard methods based on co-occurrences networks and supervised machine learning algorithms. In order to compare such assessments, we develop an out-of-sample prediction framework based on the assumption that, on average, the next technology sector in which a firm will patent will be among the ones that are more related with its present patenting portfolio. We find that machine learning algorithms not only show better prediction performances but allow for a two-dimensional representation of technology sectors that we call Continuous Technology Space (CTS). The CTS can be used to visualize the patenting portfolio of companies and to design strategic investments and acquisitions.

\section*{Results}
The data we will use in this study is the matrix representation of the bipartite company-technology networks. In particular, we will consider 643 technological sectors embedded in the patents submitted by 197944 firms in 12 years. In practice, we will use 12 $\textbf{V}^{(y)}$ matrices that link the layer of firms with that of technology codes, where $y$ ranges from 2000 to 2011. The matrix element $V_{ft}^y$ is the number of patents submitted by firm $f$ in technology $t$ in year $y$. In the following, we will interchangeably use the terms technological code, sector, or simply technology to express the same concept, since the codes written in the patents do represent technological sectors and so, in this sense, technologies. This information is obtained by matching the \textit{AMADEUS} database (\url{https://amadeus.bvdinfo.com}), that covers over 20 million firms with European registered offices, with the \textit{Patstat} (\url{www.epo.org/searching-for-patents/business/patstat}) database about patents submissions.
More details can be found in the Methods section and in \cite{pugliese2019coherent2}. \\ The matrix element ${V}_{f,t}^{(y)}$ gives a continuous quantification of the patenting activity of firm $f$ in the technology $t$. However, in the EFC framework one usually deals with binary matrices; our choice is to use different thresholds $T$ and to define 12 $\textbf{M}^{(y)}$ matrices, one for each year from 2000 to 2011:
    \begin{equation*}
    M^y_{ft} = \begin{cases} 1\ \ \text{if}\ \ V^y_{ft} \ge T\\
0\ \ \text{if}\ \ V^y_{ft} <T.
\end{cases}
\end{equation*}
So the element $M_{f,t}^{(y)}$ is equal to $1$ if a firm $f$ submits more than $T$ patents with technological code $t$ in the year $y$, and $0$ otherwise. We point out that in the Economic Complexity framework one usually binarizes the export (or, if patents are considered, the innovation) matrix using Balassa's Revealed Comparative Advantage \citep{balassa1965trade,hidalgo2018principle,pugliese2019unfolding}. Since in this case the $\textbf{V}$ matrix is very sparse, the effect of RCA is practically negligible so we preferred to use the number of patents for clearer interpretability. \\
From these $\textbf{M}$ matrices we can train different algorithms to calculate our predictions. In order to have an out of sample forecast, we use data from $2000$ to $2009$ for the training phase and to obtain a score matrix ${\textbf{S}}^{2011}$; given the matrix elements ${S}_{f,t}^{2011}$, we expect that a higher value is connected to a higher probability for firm $f$ to patent in technology code $t$ in year $2011$.
\begin{figure}[ht] 
	\centering
	\includegraphics[width=0.8\textwidth]{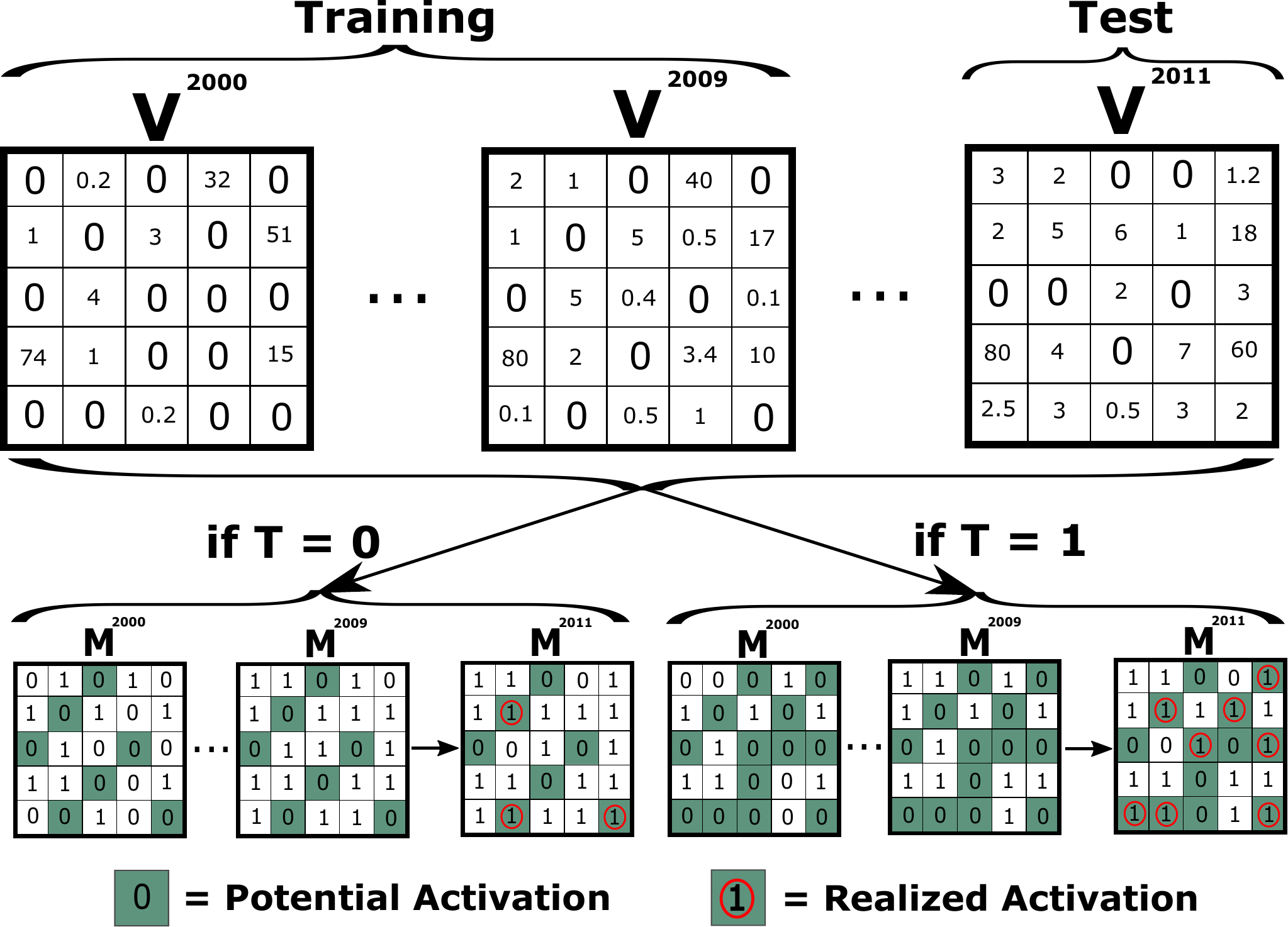}
	\caption{Schematic representation of the data processing. The \textbf{V} matrices represent the yearly bipartite networks that link firms and technologies; each element $V_{f,t}$ represents how much a technology $t$ is present in the patents of firm $f$. The elements $M_{f,t}$ of \textbf{M} matrices instead, gives us information about \textit{whether} a technology $t$ is made by a firm $f$ (or not), that is if $V_{f,t}$ exceeding (or not) a threshold $T$. The dark green elements are the potential activations and the red circles in 2011 matrices are the realized activations (i.e. those elements that were $0$ in all training years and then, in 2011, are turn on: a new technology for that firm). Also the distinction between training and test set is shown.}
	\label{fig:input_matr}
\end{figure}
\\
We point out that both the matrices $\textbf{V}$ and $\textbf{M}$ are highly \textit{autocorrelated} in time: if a firm does submit patents with a given technological code in a year $y$, it is likely that it will also in the year $y+\delta$, and viceversa. As a consequence, we focus our attention on those matrix elements that we call \textit{potential activations}: the elements of $\textbf{M}$ that  are 0 in all training years (from $2000$ to $2009$). Then, we will check whether in the test year (2011) this element remains equal to 0 or becomes 1. We will call this last case \textit{realized activation}: a firm enters, that is, starts patenting in a technological sector which is new to this firm. In Figure \ref{fig:input_matr}) we represent how we managed the $\textbf{V}$ and the $\textbf{M}$ matrices, the division of the data in training and test set, and the definitions of both potential and realized activations.\\
Our forecast exercise permits to compare different prediction algorithms using the test year 2011. So we will compute one score matrix ${\textbf{S}}^{2011}$ for each algorithm and we will compare it with $\textbf{M}^{2011}$ (obtained by binaring the empirical $\textbf{V}^{2011}$), and quantifying the prediction performance as in usual supervised classification tasks \citep{kotsiantis2007supervised}.\\
In order to obtain the scores, we use different algorithms to evaluate a relatedness \citep{hidalgo2018principle} between a firm and a technology. In the case of co-occurrences based networks, an intermediate step is to assess the \textbf{similarity} between technology codes. Here we list the tested algorithms by category, leaving a more detailed discussion for the Methods section.
\begin{itemize}
    \item \textbf{Benchmarks}: We use a quasi-trivial Random and Autocorrelation-based predictions as benchmarks.  The first is a random model where we keep the diversification of the firms $d_f =\sum_t M_{f,t}$, i.e. the number of the technology codes in its patents. The second is a benchmark model that takes into account the autocorrelation in years between the $\textbf{M}$ matrices: the scores $\textbf{S}$ are equal to the number of patents $\textbf{V}$ in the last training year.
    \item \textbf{Networks}: The standard Economic Complexity approach usually starts from the evaluation of normalized co-occurrences; in the simplest case
\begin{equation*}
    B_{t,t'}^{(y)} = \sum_f{M_{f,t}^{(y)}M_{f,t'}^{(y)}}.
\end{equation*}
Different normalizations lead to the Product Space, or in this case, the Technology Space \citep{hidalgo2007product}, the Taxonomy Network \citep{zaccaria2014taxonomy}, and the Micro-Partial network, based on the paper of \cite{teece1994understanding}. In all these cases, the network $\textbf{B}$ represents a projection of the bipartite network $\textbf{M}$ into the space of technology codes, and each element $B_{t,t'}$ represent the proximity between the two technology codes.
In order to obtain a measure of the relatedness between a firm $f$ and a target technology $t$, to be used as a prediction score, one then computes the coherence \citep{pugliese2019coherent2} using eq. \ref{eq:scores}. Other approaches, such as the density normalization introduced by \cite{hidalgo2007product}, perform sensibly worse.
\item \textbf{Machine Learning}: Since our prediction exercise can be expressed in a supervised classification exercise, we can use the Random Forest algorithm \citep{breiman2001random,albora2021product}, and what we call the Continuous Technology Space (CTS). The first is a popular machine learing algorithm based on decision trees, while the CTS is based on the studies of  \cite{tacchella2021relatedness}, and it is a projection on the space of the technology codes of the scores obtained with the Random Forest. This is done by using a Variational Auto Encoder \citep{kingma2013auto} followed by the t-SNE dimensional reduction algorithm \citep{van2008visualizing}. In this way we are able to make the results of the Random Forests more interpretable. Even if the CTS is based on Machine Learning, in order to produce prediction scores one has to compute a coherence or density measure as in the network based approaches.\\
Two type of Random Forest are used, the non-Cross Validated (RF) and the Cross Validated one (RF\_CV). With the cross validation, we remove a portion of firms at a time from the training, and then we use them in the test. The idea is that the algorithm produces its predictions by using two pieces of information: the similarity between technologies and its ability to recognize a firm. By cross-validating the RF we try to force the algorithm to use the former, and not the latter \citep{albora2021product}.\\
\end{itemize}

\subsection*{Prediction results}
\begin{figure}
\centering
\subfloat
   {\includegraphics[width=.45\textwidth]{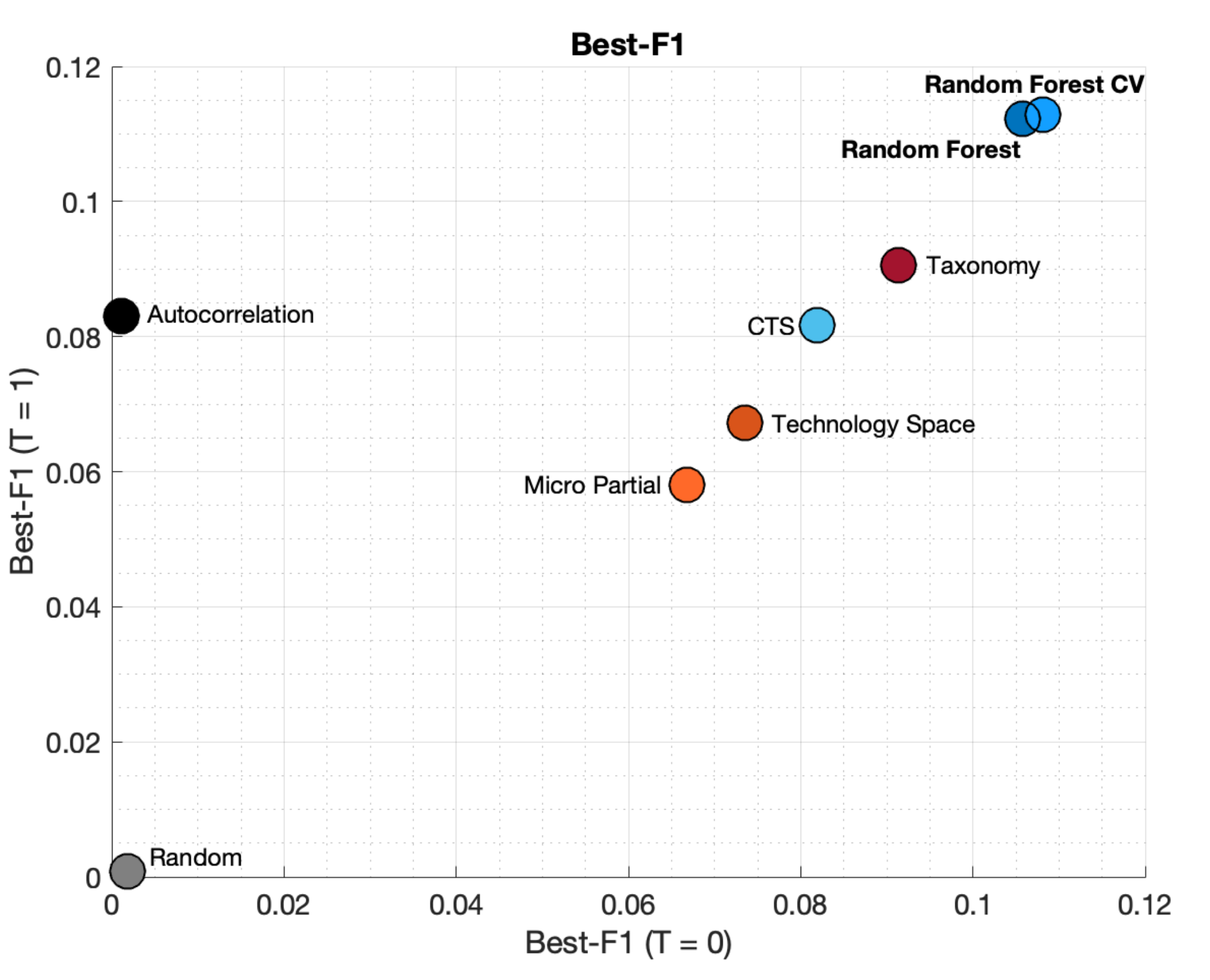}\label{fig:BF11}} \\
\subfloat
   {\includegraphics[width=.45\textwidth]{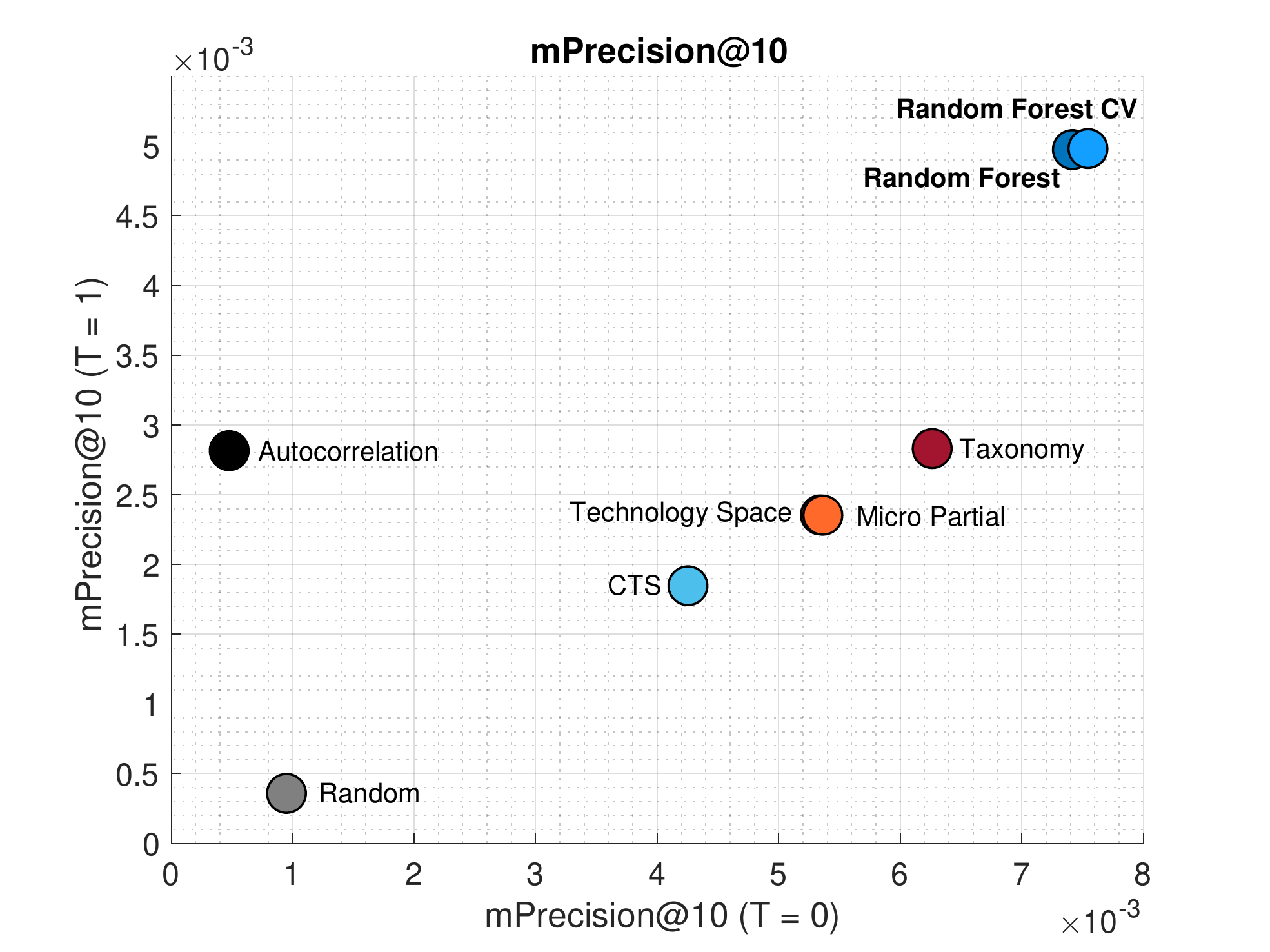}} \quad
\subfloat
   {\includegraphics[width=.45\textwidth]{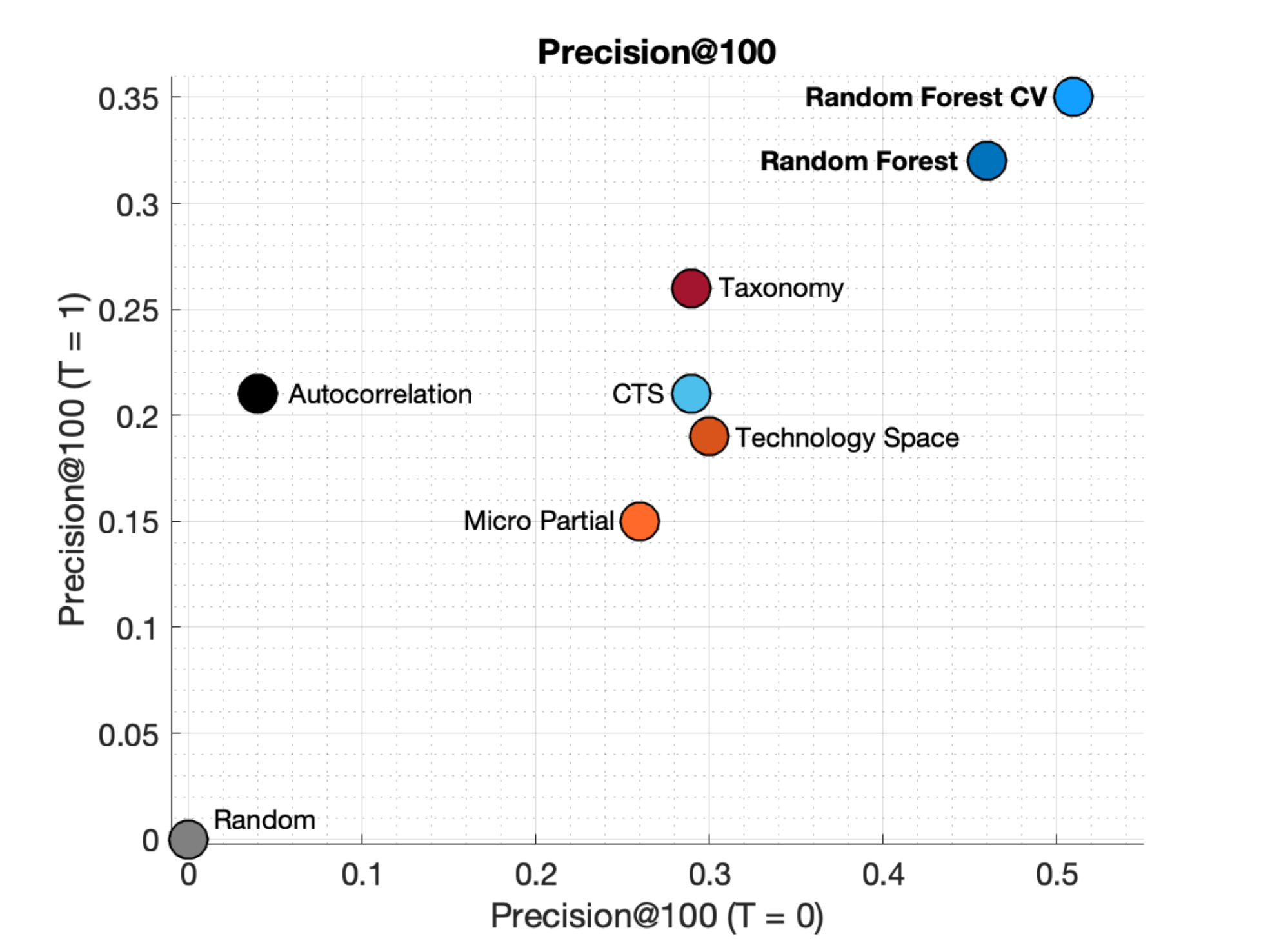}}
\caption{Comparison of the prediction results obtained by different approaches and using three different evaluation metrics and two binarization thresholds. Random Forest over-perform all other approaches in all perspectives. Regarding the Best F1 and the Precision@100 metrics, the second bests are the Taxonomy Network and the Continuous Technological Space.}
\label{fig:results}
\end{figure}

\begin{figure}
    \centering
    \includegraphics[scale=.55]{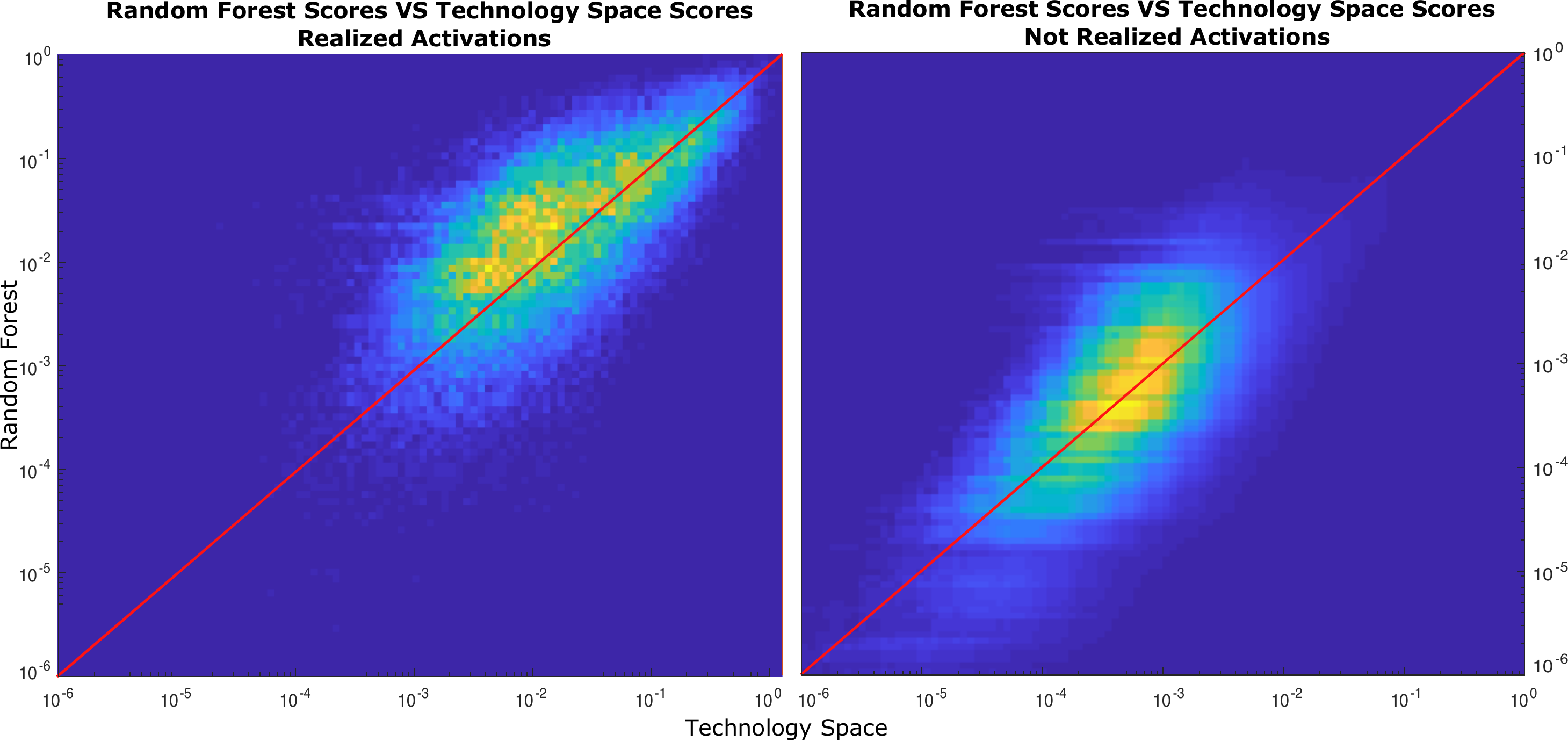}
\caption{Comparison of the Random Forest and Technology Space scores. The left figure is referred to the Realized Activations, i.e. those elements that are always 0 during the training years (from 2000 to 2009) and then become 1 in the test year (2011). The red line is the bisector. The scores obtained with the Random Forest are, on average, higher than those obtained with the Technology Space. The right figure is referred to the Not Realized Activations, i.e. elements that also in the 2011 are 0s. Here the distributions are roughly similar.}
\label{fig:RF_VS_TS}
\end{figure}

Here we compare the co-occurrences based networks with the machine learning algorithms, showing how the latter are able to give better prediction results. The results are shown in Figure \ref{fig:results}.
In order to compare the various prediction methods from various viewpoints, we adopted different metrics to quantify the goodness of a predictions (these metrics are better presented in Methods section):
\begin{itemize}
    \item \textbf{Best-F1}: the harmonic mean of Precision and Recall, maximized in-sample in order to minimize both false positives and false positives;
    \item \textbf{Precision@100}: the fraction of the first 100 elements of the score matrix $\textbf{S}^{2011}$ that are actually activated;
    \item \textbf{mPrecision@10}: we consider the first 10 scores for each firm and we compute the fraction of realized activations; then we average over the firms.
\end{itemize}
In those Figures we report the scores of the previous metrics for different values of the threshold parameter $T$; the results are consistent even if one varies such threshold (or uses the RCA to binarize).\\
We start noticing that the random benchmark is surpassed by all the different approaches, showing that all are able to compute a measure of similarity that is able to grasp the some links between the technology codes.\\
However, the autocorrelation benchmark performs better than most density based measures, including the CTS. In particular, it
performs better when T increases, because the number of zeros in both the training and the test matrices increases (that is, the number of potential activations, the green elements in Figure \ref{fig:input_matr}, that are not realized).\\
In the Best-F1 and Precision@100, the only network-based algorithm that manages to overcome the CTS is the Taxonomy. In particular, it is interesting to observe how this Network exceeds the Technology Space. We can argue that for the technology codes, a network based on taxonomy, i.e. how firms move from low-complexity to high-complexity technologies only after developing the necessary skills \cite{zaccaria2014taxonomy}, shows a better prediction performance that a proximity one, i.e. a network where two technologies have an high link if they have the same capabilities \citep{hidalgo2007product}.\\
The Micro Partial approach does not show a competitive performance despite being quite popular in both academic and corporate applications \citep{smith2017two}.\\
In any case, the superiority of the RF\_CV and of the RF with respect to both benchmarks and density-based approaches is evident. Although the other algorithms are able to give prediction scores able to overcome the benchmark models (especially true for $T = 0$), clearly these are not able to fully highlight the non-linear relationships among the technological portfolios of firms and the technological sectors they will move to.\\
Finally, in Figure \ref{fig:RF_VS_TS} we compare the distribution of the scores of both the realized and the not realized activations for the Random Forest and Technology Space. In order to make them comparable, both scores are rescaled using the respective maxima and minima. The red lines is the bisector, showed for further reference. From the left figure it emerges that the Random Forest assigns, on average, higher scores to those activations which will be actually realized in two years. On the contrary, the possible but not realized activations show similar distributions; this is due to the much greater number of true negatives which is present in both approaches. 
Note that, as expected, the scores given to the Not Realized Activations are lower than the Realized ones.
\subsection*{Continuous Technology Space}
\begin{figure}
    \centering
    \includegraphics[scale=0.85]{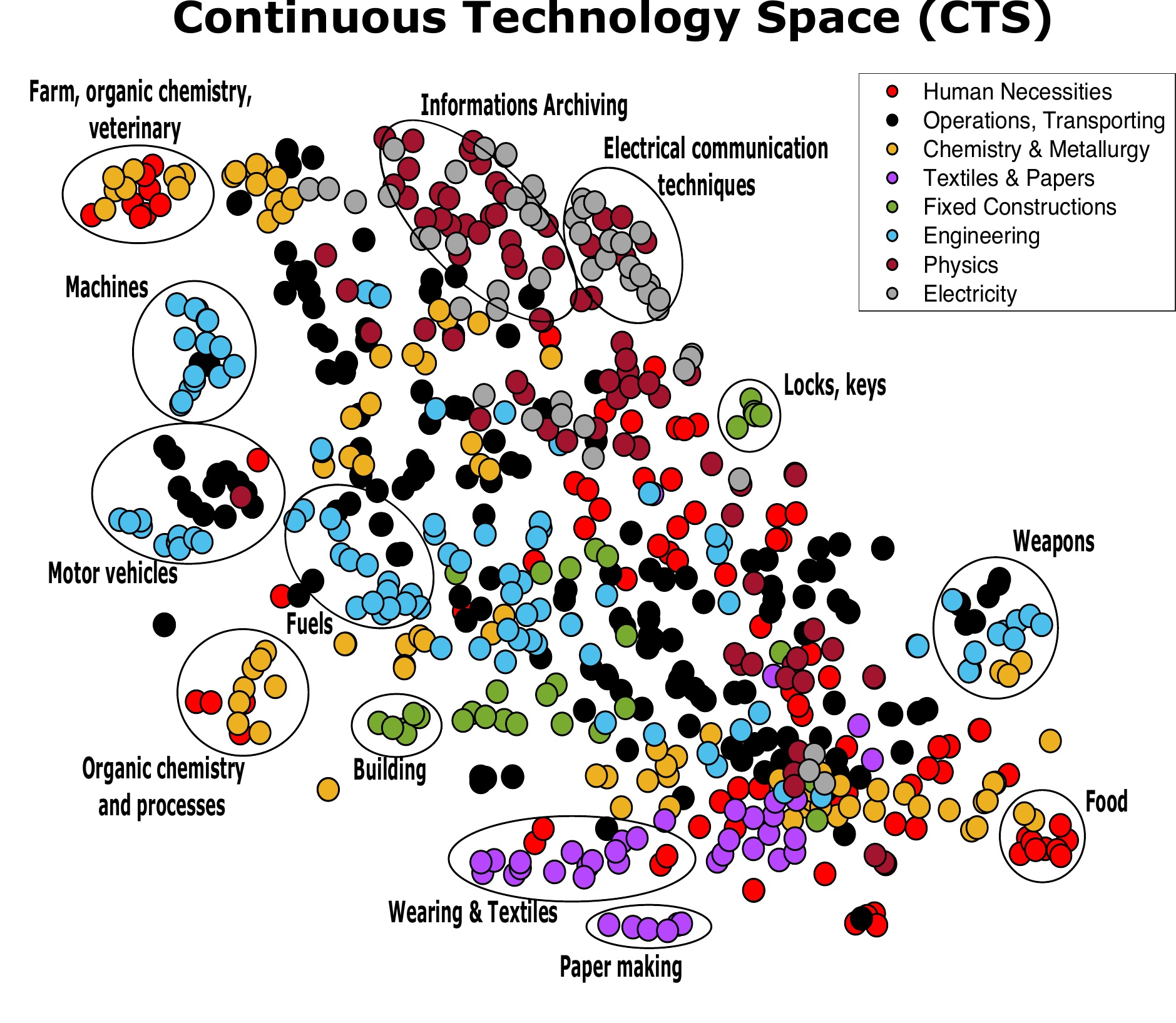}
\caption{Representation of the Continuous Technology Space (CTS). Each point is a technology code and each colour is associated to the respective macrocategory, i.e. the first of the 4 digits. Clusters related to macro categories are evident.}
\label{fig:CTS}
\end{figure}

\begin{figure}
    \centering
    \includegraphics[scale=0.65]{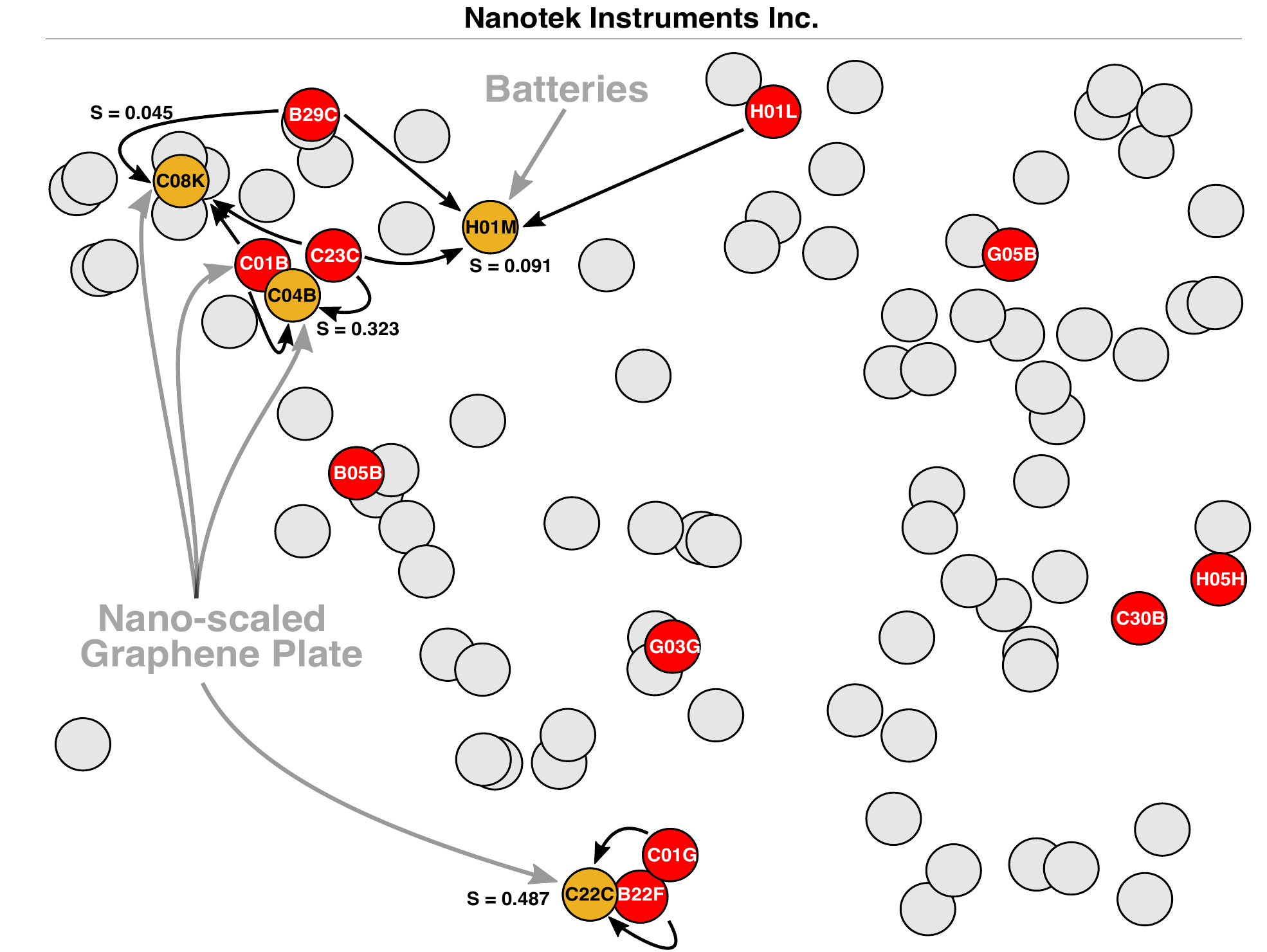}
\caption{The exploration of the Continuous Technological Space by an American nanotechnology company. Starting from the red sectors, in which Nanotek patented in the past, this company moved nearby, patenting in the gold sectors in the next years.}
\label{fig:CTS_N}
\end{figure}

Even if the prediction performance of Random Forests vastly overperforms the other approaches, their practical feasibility is limited by their low interpretability. From a policy perspective, indeed, it is not easy to justify a strategic decision such as to invest or not in a technological sector on the basis of a quasi-black box algorithm. For this reasons we introduced the Continuous Projection Space \citep{tacchella2021relatedness}, that uses the scores obtained from the machine learning algorithms to build a two-dimensional, easy interpretable space to describe the temporal evolution of bipartite networks. Here we apply this methodology - which is fully described in the Methods section - to the firm-technology network, and we discuss an exemplary application. In Figure \ref{fig:CTS}) each point represents a technology code, and the different colors correspond to IPC macro-categories, i.e. the first of the 4 digits of the classification codes. We point out that, differently from network-based representations, here the similarities are simply represented by the spatial proximity between technology codes. The use of euclidean distances instead of topological ones permits to use a wider range of tools, for instance clustering and anomaly detection algorithms. A visual inspection of the CTS permits to obtain a number of insights: in the Figure \ref{fig:CTS}) can be observed that technology codes tend to cluster corresponding to macro categories; therefore the positions of the technological codes on the plane obtained with t-SNE are not random but present a certain degree of significance. In the CTS, we can observe the presence of some dense parts where it is possible to find veterinary medicine close to farm; in the Motor vehicles area we find motor vehicle technology codes: in particular here there is a red colour technology code (A47C) that corresponds to chairs and seats specially adapted for vehicles, black technology codes color, corresponding to B60 (considering the first 3 digits), that represent vehicles, light blue technology codes corresponding to the first 3 digits F01 and F02, i.e. machines and engines, and combustion engineering, and a brown technology code color (G05G), physics of command systems; Weapons area is associated with weapons technologies: we find principally (considering the first 3 digits) codes B63 and B64, i.e. ships and aircrafts, C06 associated to explosive chemistry, and F41 and F42, i.e. weapons and ammunition.\\
With, as examples, Motor vehicles and Weapons areas it is possible to observe how, although sometimes codes belonging to different macro categories are mixed, the CTS does this following a certain coherence.
\\
In order to show a practical application of the CTS, we show in Figure \ref{fig:CTS_N} the portion of this space relative to an American nanotechology company, Nanotek Instruments Inc., as an example.
In 2002 Nanotek patented three inventions, two based on batteries (\url{https://patentimages.storage.googleapis.com/f4/d8/3d/d663e43fe48e2b/US6773842.pdf} and \url{https://patentimages.storage.googleapis.com/66/b3/7f/6fa873ae402fbf/US6864018.pdf}) and the third is the \textit{Nano-scaled graphene plates} (\url{https://patentimages.storage.googleapis.com/e5/3d/0d/1c25e5f68a77ab/US7071258.pdf}). The first two are associated to the code \textit{H01M}, while the third to the codes \textit{C08K}, \textit{C04B}, \textit{C01B} and \textit{C22C}, that correspond to the gold points. The red points are the technology codes in which Nanotek patented in 2000 and 2001, while, as mentioned, the gold ones are that activated in the 2002. The black arrows underline the non random position of the new technologies, that are found close to the ones already present in the patenting activity of the company. This is because we find that technology codes that have a high similarity are represented close to each other, and therefore a sort of "technological diffusion" is expected starting from the codes that firms already have in their portfolios.\\

\section*{Discussion}
In this work we compare machine learning and network-based approaches to forecast which will be the future patenting activity of firms; in particular, their next technological sector of innovation. To the best of our knowledge, this is the first attempt to assess the relatedness between a firm and a technology sector using machine learning. In order to compare the various possible measures of relatedness we analyze a very large database consisting in about two hundred thousand firms and 643 technology sectors and we develop a forecast exercise using the assumption that, on average, firms will patent in sectors related to their present technological activity. We find that supervised machine learning techniques (Random Forest) clearly overperform the standard methodologies usually adopted in Economic Complexity, that is, networks of co-occurrences. Our results are robust with respect to different definitions of what a ``new'' technological sectors is, and if different metrics to evaluate the prediction performance are adopted. Indeed, Random Forest assigns on average higher activation scores to those technologies which will be explored by firms with respect to all network based approaches. Finally, we introduce the Continuous Technology Space (CTS), that permits to visualize the dynamics of firms during their innovation activity.
The introduction of this approach opens a number of possible applications and developments. First of all, our activation scores represent an assessment of the achievability of a given jump to a new technology sector, a measure of how easy will be to produce innovations in that sector given the present activity of the firm. Moreover, the CTS allows a compact representation of the past, the present and the possible patenting activity of a firm: using this tool it is possible to quantify how much a firm is \textit{exploring} the space of technologies or \textit{exploiting} what it already does. One can then compare the strategy with various measures of performance, both in terms of profitability and further innovation activity. Finally, these measures can be applied to investigate Mergers and Acquisitions, and in particular to study whether acquirers prefer to target companies which are ``close'' or ``far'' from their present patenting activity.

\section*{Methods}
In this Section we describe in more detail the database, algorithms and metrics used in the analysis.
\subsection*{Data}
The bipartite firm-technology network is obtained by matching two database: AMADEUS for firms and PATSTAT for the technology codes.
\subsubsection*{Firms}
AMADEUS (\url{https://amadeus.bvdinfo.com}) contains information about over 20 million companies, mainly concentrated in the European continent. This database is managed by Bureau van Dijk Electronic Publishing (BvD) which specializes in providing financial, administrative and budget information relating to companies. It is compatible with the PATSTAT database for patents as BvD includes the same patent identifiers as the European Patent Office \citep{pugliese2019coherent2}. We mention one of the well-known problems with AMADEUS is that large companies are fully covered while those with less than 20 employees are under-represented \citep{ribeiro2010oecd}; this is not a severe issue for the present analysis.
\subsubsection*{Technology Codes}
The dataset from which we take information about the patent and the technology codes is PATSTAT (\url{www.epo.org/searching-for-patents/business/patstat}). This database contains information about approximately 100 millions of patents registered in approximately 100 Patent Offices. These information spans from mid-19th century to three-four years before release of the database; this is evident from the quickly decreasing number of patents in the last available years. As a consequence, we decided to restrict our analysis in a conservative time interval. A key element is the present of a set of alphanumeric codes in each patent submission; these codes can be assigned by the inventors or by the reviewers and represent the technological sector the patent belongs to. The WIPO (World International Patent Office) uses the IPC (International Patent Classification) \citep{fall2003automated} to assign these technology codes to each patent in such a way as to classify, and better manage, the inventions presented. The IPC codes define a hierarchical classification consisting of six levels: sections (that we call macro category), sub-sections, classes, sub-classes, groups, sub-groups. For example, code Axxxxx corresponds to the "Human Necessities" macro category and Hxxxxx to the "Electricity" macro category; considering the following digits we have, for example, with A01xxx the sector "Agriculture; Hunting", and with A43xxx the "Footwear" sector. It is important to note that we discard classes ?99? and sub-classes ?Z?, as they represent other technologies not classified in other classes or sub-classes, and they are therefore not well defined.\\
It may happen that the same invention may be referred to for multiple patent application documents. In this case, each group of documents in PATSTAT is called "Patent Family" according to primary citations among them \citep{oecd2001oecd}, which is nothing more than the set of patents presented in different countries to protect the single invention. Patent Families can be built with different criteria \citep{martinez2011patent}, but among these we choose the one related to the "Extended Family", also called IN-PADOC. This corresponds to the category considered in such a way as to associate the inventions with the widest possible technological spectrum. Once patents are assigned to firms, we can assign them the corresponding technology codes and build the firm-technology bipartite network, and its adjacency matrix $\mathbf{V^{(y)}}$, one for each year $y$. The interested reader can find more details about this data in the work of \cite{pugliese2019coherent2}.

\subsection*{Data processing}
The starting database can be represented using the following structure: $12$ matrices, one for each year from $2000$ to $2011$, that link $426983$ firms $f$ (rows) to $7456$ (6-digits) technology codes $t$ (columns). We chose to work at a higher aggregation level, and so to compress the technology codes from 6 to 4 digits, summing the columns corresponding to the 6 digit codes with the same first 4 digits. From the 6 to the 4 digit level the number of technologies goes from $7456$ to $643$. This operation leads to both better quantitative results and shorter computation times (from a qualitative point of view, instead, the results are unchanged).\\
A key element of both the machine learning and the network based approaches is to provide an assessment of the \textbf{similarity} between technology codes; this information can be extracted from the co-occurrences of technological sectors in the same firms. So we consider only firms that, in years from $2000$ to $2009$, make at least $2$ technology codes; these firms are $197944$.\\
This leads to the data mentioned to the main text: 12 $\textbf{V}$ yearly matrices that link $197944$ firms and $643$ technology codes.\\
In order to compute the relatedness measures, in the Economic Complexity literature \citep{hidalgo2007product,zaccaria2014taxonomy,pugliese2019unfolding} one usually computes the Revealed Comparative Advantage or RCA \citep{balassa1965trade}, and then these matrices are binarized using a threshold equal to $1$. As far as exports are concerned this choice of threshold has a natural economic meaning, traceable to the works of Ricardo and Balassa himself: considering the bipartite country-product network \citep{tacchella2012new}, $RCA_{c,p} \ge 1$ means that country $c$ is significantly competitive in the export of the product $p$; so the country's share of that product in its market is equal to or greater than the product's share on the world market. However, the economical meaning of patents submission is different, so the choice of RCA is not straightforward. In this work, we binarize the matrices $\textbf{V}$ with different values of threshold $T$, without computing the RCA; in this way, the matrices $\textbf{V}$ are better interpretable as a firm presenting more than $T$ patents with the technology code $t$. We have in any case checked the robustness of our results for different threshold values and the use of RCA.

\subsection*{Network-based approaches}
In this and in the next sections we discuss how to obtain a prediction score matrix \textbf{S} for 2011 from each method starting from the same training data $\textbf{V}$ and $\textbf{M}$, relative to the years 2000-2009. The score matrix gives the model's estimation of the likelihood that a firm will patent in the given technology sector, and the comparison between the scores and the actual $\textbf{M}^{2011}$ using the performance metrics will give an assessment of the models' performance.\\
The basic idea of network-based approaches is to compute a similarity of technology codes from their co-occurrences in companies. Introduced by \cite{teece1994understanding}, and popularized in the network/complexity community by \cite{hidalgo2007product}, the basic quantity the number of companies that have patented inventions relating to both codes:
\begin{equation*}
    B^{CO}_{t,t'} = \sum_{f}{M_{f,t}M_{f,t'}}.
\end{equation*}
The idea is that if many firms are active in two technology sectors $t$ and $t'$ at the same time, this means that the capabilities, the techniques and, in general, the necessary means to patent in these sectors, are roughly the same, and so these sectors are, in this sense, similar, or related.\\
Different scholars presented various ways to normalize the co-occurrences, on the basis of different theoretical frameworks or interpretations. In general, we can write:
\begin{equation*}
    B_{t,t'} = \frac{1}{A}\sum_{f}{\frac{M_{f,t}M_{f,t'}}{C}}
\end{equation*}
and discuss the various options for the quantities $A$ and $C$:
\begin{itemize}
    \item Simple Co-Occurrences \citep{teece1994understanding}: for $A = 1$ and $C = 1$ one simply counts the number of companies that are active in both sectors;
    \item Technology Space (same normalization of the Product Space \citep{hidalgo2007product}): $A = \max{(u_{t},u_{t'})}$ and $C = 1$, where $u_t = \sum_{f}{M_{f,t}}$ is the \begin{itshape}ubiquity\end{itshape} of technology code $t$, that is, the number of firms active in that technology sector. Using this type of normalization we give a lower connection weight to those technology codes done by many firms, that we can consider as basic.
    \item Taxonomy \citep{zaccaria2014taxonomy}: $A = \max{(u_{t},u_{t'})}$ and $C = d_f$, where $d_f = \sum_{t}{M_{f,t}}$ is the \textit{diversification} of firm $f$. The Technology Space, for how it is built, gives a higher score for high complexity technology codes (i.e. codes done by few firms) and, as a result, bias towards them. Consequently, it is not possible to justify the evolution of low-complexity technology codes towards high-complexity ones. Normalizing also for the diversification we avoid this problem as we penalize low ubiquity scores and low complexity technology codes are weighed more.
    \item Micro Partial \citep{teece1994understanding}: we compute
    \begin{equation*}
        B^{MP}_{t,t'} = \frac{B^{CO}_{tt'}-\mu_{tt'}}{\sigma_{tt'}}
    \end{equation*}
    with
    \begin{equation*}
        \mu_{tt'} = \frac{u_tu_{t'}}{N},
    \end{equation*}
    and
    \begin{equation*}
        \sigma^2_{tt'} = \mu_{tt'}\frac{(N-u_t)(N-u_{t'})}{N(N-1)},
    \end{equation*}
    where $N$ is the number of companies. Here we use a null model in which the ubiquities of the technologies are kept fixed and everything else is randomized. This case can be analytically solved: the resulting distribution for the co-occurrences is hypergeometric with mean $\mu_{tt'}$ and variance $\sigma^2_{tt'}$. We call this network Micro Partial following the notation used by \cite{cimini2021meta}: this null model is microcanonical because the degree sequence is exactly fixed and partial because only one layer is constrained. So the idea is that, if the weight of the link between two technology codes $t$ and $t'$ exceeds the expected value $\mu_{tt'}$, this means that $t$ and $t'$ are highly related with respect to this random case. Furthermore, as a t-statistic, $B^{MP}_{t,t'}$ measures how much the observed link between the two technology codes exceeds what would be expected if the companies were randomly assigned.
\end{itemize}
For the latest formulas we obtain one matrix $\textbf{B}^{Net}$ for each network. In order to consider all years available in the training data, we using as $\textbf{M}$ matrix in the previous formulas a total matrix obtained by summing the $\textbf{V}$ matrices from years $2000$ to $2009$, using all the $197944$ firms, and then binarizing this sum.\\
Based on the network used, we get a $\textbf{B}^{Net}$ which we use in the coherence equation from \cite{pugliese2019coherent2}:
\begin{equation}
    {S}_{f,t}^{2011} = \sum_{t'}M_{f,t'}^{2009}B^{Net}_{t't},
\label{eq:scores}
\end{equation}
where $M_{f,t'}^{2009}$ is the $\textbf{M}$ matrix obtained by binarizing the $\textbf{V}^{2009}$ matrix. In practice, $t$ is highly coherent with the patenting activity of firm $f$ if $f$ is active in many sectors highly connected with $t$. On the contrary, if a sector is far from what a firm actually does, we will assign to it a lower activation score. Note that this equation differs from the density equation of the Product Space \citep{hidalgo2007product}; we use coherence instead of density since we have found a better predictive performance.

\subsection*{Random Forest}
\begin{figure}
    \centering
    \includegraphics[scale=0.69]{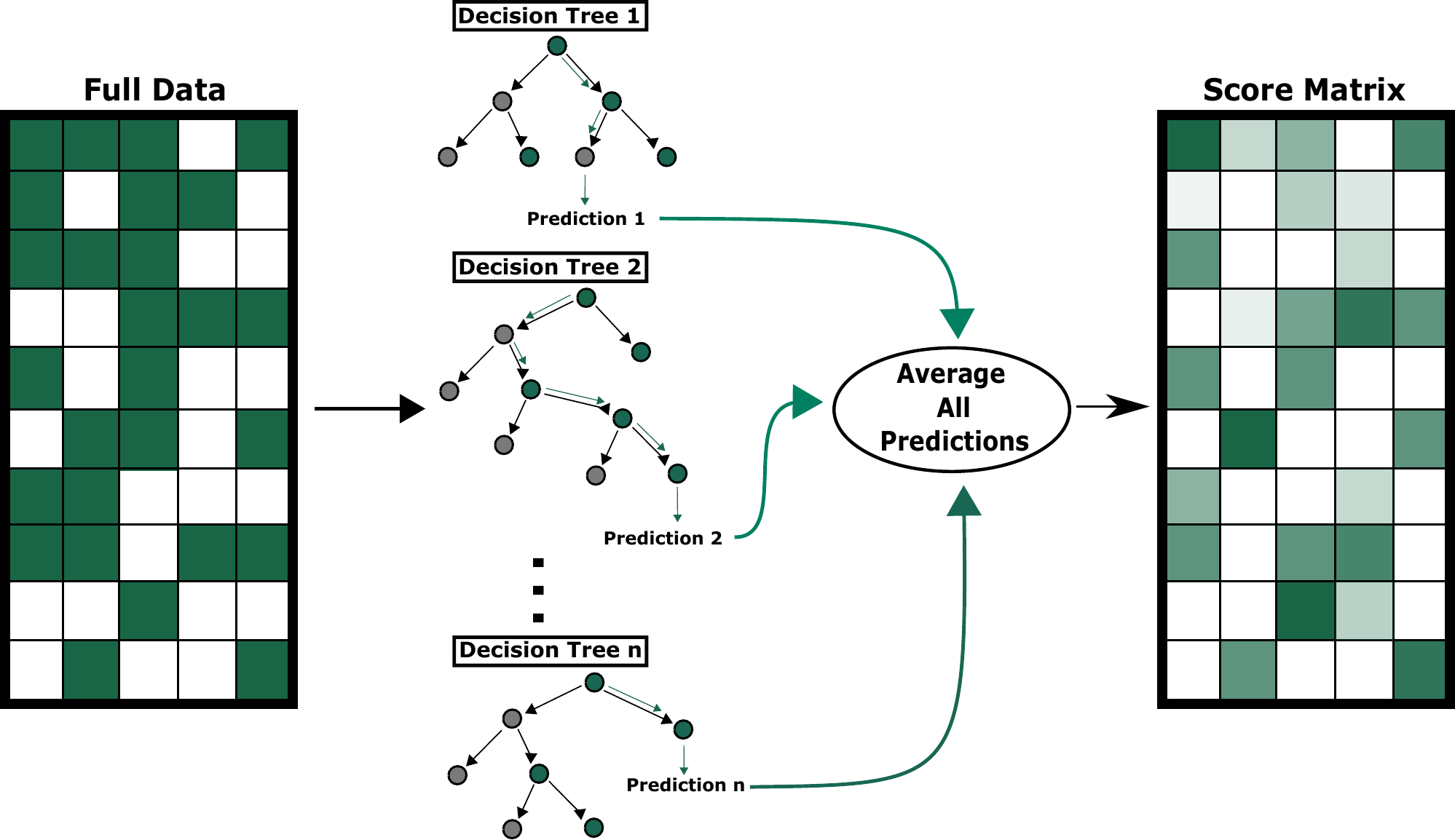}
\caption{We show how the score matrix $\textbf{S}^{2011}$ is obtained by the Random Forest by combining the predictions of different decision trees.}
\label{fig:met1}
\end{figure}

Random Forest \citep{breiman2001random} (RF) is a tree-based machine learning algorithm that we use to better capture the non-linear links between technology codes. In particular, we use this binary classification algorithm to determine whether or not a technology code will appear in the patenting portfolio of a particular company in the future starting from the knowledge of the technology codes in which the firms patented in the last training year.\\
In general, during the training of a supervised machine learning algorithm an input data $\textbf{X}$ matrix is passed; each vector of the matrix is a sample with a number of features equal to the shape of the vector, which must be associated with the labels which, in classification problems, are present in a different input $\textbf{y}$. To give an example, $\textbf{X}$ can be the matrix where each row is a flattened handwritten digit, and each element of the row is the intensity of a pixel; in this case $\textbf{y}$ will be the label corresponding to the digit, and that must be associated, in order to be recognized, to all those present in $\textbf{X}$. Once the model is trained, one gives new samples $\textbf{X}_{test}$ and the model is able to make associate a prediction $\textbf{y}_{test}$(in this case, a digit), to each sample.\\
In our case, we train one RF for each technology code: we want the RF to learn to which typologies of portfolios is associated each code after two years. So, as samples matrix $\textbf{X}$ we use the matrix obtained by concatenating, or stacking vertically, the $\textbf{V}$ matrices from the year $2000$ to $2007$, and as $\textbf{y}$ we use one column at a time (and therefore one technology code at a time) of the matrix obtained by concatenating the matrices $\textbf{M}$ from the year $2002$ to $2009$. In this way, each row is a firm in a year from $2000$ to $2007$, and the RF has $643$ features. We associate this row to the respective label in $\textbf{y}$, that is, if after $2$ years the technology code associated to the element in $\textbf{y}$ is active, or not. In such a way, we associate the codes of each patents to the possible presence of the target code in the future.\\
From a practical viewpoint, we use the "RandomForestClassifier" from the "sklearn.ensemble" python library \citep{scikit-learn}, called in this way: 
\begin{equation*}
    \text{RandomForestClassifier.fit}(\textbf{V}_{2000}^{2007}, \overrightarrow{\textbf{M}}_{2002}^{2009}),
\label{RF1}
\end{equation*}
where $\textbf{V}_{2000}^{2007}$ are the vertically stacked matrices and with the vector symbol over $\textbf{M}$ we indicate that one column is used at a time, that is, we train one RF for each technology code. The delay of 2 years is used to insert a dependence on time, as we want to produce forecasts about the innovative development of firms.
We optimized the RF parameters as described in the supplementary information; the results show here refer to: number of trees $= 50$, min$\_$samples$\_$leaf = 4; max$\_$depth = 20 and method = 'entropy'. The use of all available companies in the training is computationally demanding, so we used only the top 10000 most diversified firms (10KHD firms). If we use more firms, we practically obtain the same results because, as the number of firms used for training increases, we get a saturation of the forecast performances (see the Supplementary information). The fact that firms with higher diversification should be used is due to the fact that these provide a better coverage of the possible technologies and the possible combinations among them.\\
After fitting the data, that is, training the model, we obtain the ${\textbf{S}}^{2011}$ scores by using the $\textbf{V}^{2009}$ matrix as $\textbf{X}_{test}$. The command line reads
\begin{equation*}
    {\textbf{S}}^{2011} = \text{RandomForestClassifier.predict\_proba}(\textbf{V}^{2009})
\label{RF2}
\end{equation*}
and this associates a probability to activate the target technology to each firm in 2009.
In Figure \ref{fig:met1} we schematically represent how $\textbf{S}^{2011}$ is obtained from the RF, seen as a set of differently trained decision trees.\\
In this work we compared this approach with a cross validated RF, for which we use the same parameters and the same training and test sets. The difference is the following. We train $k=4$ different RFs, using the technique called \textit{k-fold Cross Validation}. In each RF we remove from the training 1/4 of the 10KHD firms and then we use these in the test together with 1/4 of the remaining low diversification companies that are present only in the test, as in the previous training. To be more specific, the $\textbf{X}_{test}$ should always be consistent with the $\textbf{V}^{2009}$ that has all the 197944 firms. Of these, the first 10K HDs are used as training, so we split four times these 10K HD firms and the remaining ones, and for four times the training is done without the 1/4 of the 10KHD firms, and the test is done using these left out firms and 1/4 of all the remaining low diversified firms.\\
The idea behind the use of cross validation is the following. During the training the RF basically learns two pieces of information: to recognize the portfolio of a company and the similarity among technologies. Even if we are more interested to the latter, the learning of the two can not be avoided. However, we can try to force the algorithm to use the similarities in the test phase: if we give a new company in the $\textbf{X}_{test}$, the RF can not recognized it and so it is forced to use the similarities to produce its predictions. This procedure, even if computationally more demanding, leads to better results, as shown in Figure \ref{fig:results}.

\subsection*{Continuous Technology Space}
Random Forest shares with most of the machine learning algorithms an intrinsic difficulty of interpretation, i.e. the rationale behind how the input is connected to the output is not evident. In this respect, network approaches (note: if suitably filtered) are more clear, since the coherence or density based approach are clearly visualizable: a technology is coherent with a firm's portfolio if has a lot of heavy connections with what the firm already does. In order to restore the interpretability of networks and keep the predictive performance of machine learning,  \cite{tacchella2021relatedness} propose the Continuous Projection Space, that here we reformulate, with suitable modifications, as the Continuous Technology Space (CTS).\\
To compute the CTS we starting from Random Forest CV method but, as $\textbf{X}$, the first 2K HD firms are used because of the same reason of the Random Forest method: we have a saturation of the scores, so using more firms doesn't change the scores and increase the computational time.\\
Another difference with the Random Forest CV is that the previsions are obtained using as $\textbf{X}_{test}$ the same 2K HD firms in such a way as not to hard the dimensional reduction process described below.
At the end we obtain a scores matrix of shape $[N\times\text{years}]\times[\text{\#t}]$, where $N$ in the number of companies ($N = 2000$), years $= 10$ and $\#t =$ number of technology codes $= 643$; in total this scores matrix has shape $20000\times643$.\\
Each column of the score matrix represents the likelihood that each company (rows) will patent in each technology code (columns). We can then argue than two sectors are \textit{similar} if the RF predicts that the same companies will or will not produce patents in these sectors. In this sense, the columns of the score matrix
can be seen as the coordinates in a high-dimensional space for each technology codes, where the number of dimensions is given by the number of companies multiplied by the number of training years. Obviously, it is impossible to visualize this continuous space of technologies in such an high dimensionality; so project these points in a lower dimension by combining a Variational - Autoencoder Neural Network \citep{kingma2013auto}, to reduce the dimension from $20000 \to 150$, and then t-SNE \citep{van2008visualizing}, to reduce the embedding space from $150 \to 2$ dimensions, finally obtaining the Continuous Technology Space (CTS), that we show in Figure \ref{fig:CTS}. Now the similarity between technology codes is simply given by the relative distance in this $2-D$ space, and the black-box issue is also solved, since it is easy to understand and visualize how firms move from the codes already present in their portfolios to the ones that are immediately close, as shown in Figure \ref{fig:CTS_N}.\\
Now we want to use the idea of a coherent diffusion in this low dimensional space to produce forecasts; in practice, to obtain a score matrix $\textbf{S}^{2011}$ to compare with the possible activations of 2011. We start by computing a similarity matrix for the CTS, that for the sake of simplicity we keep calling $\textbf{B}$. We use the distances between technology codes on the CTS and gaussian kernels:
\begin{equation*}
B_{i,j} = \frac{e^{-||\textbf{y}_i - \textbf{y}_j||^2/2\sigma_i^2}}{\sum_k e^{-||\textbf{y}_i - \textbf{y}_k||^2/2\sigma_i^2}},
\end{equation*}
where $\textbf{y}_i$ is the coordinate of the $i$-th technology code in the CTS (i.e. in the $2-D$ space). The $\sigma_i$ is the standard deviation of the Gaussian kernel related to the technology code $i$-th; this parameter can be set differently for each $i$ code, through a binary search process in which a quantity that quantify the number of first neighbors is fixed. As we can see in Figure \ref{fig:CTS}, there are codes in dense areas and codes in less dense areas, so the idea is to assign a high sigma value to the codes in less dense areas and low sigma values in more dense areas in order to keep the interaction with the number of first neighbors constant. The binary search process is described in the supplementary information where we also show that the best optimal value of nearest neighbors is 75.\\
After the similarity matrix $\textbf{B}$ is obtained, one can compute the score matrix $\textbf{S}^{2011}$ from the coherence Equation \ref{eq:scores}:
\begin{equation*}
    S_{f,t}^{2011} = \sum_{t'} M^{2009}_{f,t'} \cdot B_{t',t}.
\end{equation*}
In particular, this value of nearest neighbors is calculate out of sample using the 4-fold cross validation as the Random Forest CV: we use 3/4 of the companies to determinate the value of nearest neighbors that maximize the Best-F1 and than, with the remaining companies, we calculate the scores using the Equation \ref{eq:scores}.\\
In this sense the CTS is, like the network approaches, density-based: the more a firm surrounds a technology sector, the more the likelihood it will be part of its patenting activity in the near future.

\subsection*{Benchmark models}
In order understand the effective goodness of our forecast results, a comparison with some relatively trivial benchmark models is required. We used two benchmark modes:
\begin{itemize}
    \item The first is consists in a simple randomization of the technology codes. In practice, we shuffle the columns of the $\textbf{M}^{2009}$ matrix in the calculation of Equation \ref{eq:scores}. The B used is that calculated with Technology Space network starting from the not randomized $\textbf{M}^{2009}$ (using the other networks there is no significant change in the metric scores). In this way, the ubiquity of technology is kept fixed, while the diversification of firms is preserved.
    \item The second benchmark model checks the hypothesis that the simple temporal autocorrelation of the bipartite networks can explain the observed dynamics. In this case, we simply use the $\textbf{V}^{2009}$ of the test firms as score matrix $\textbf{S}^{2011}$, that is, element-wise:
\begin{equation*}
    {S}_{f,t}^{2011} = V_{f,t}^{2009}.
\end{equation*}
In this way we check if the number of patents done in the past can forecast the number of patents done in the future by the same company in the same technology sector. As shown in Figure \ref{fig:results}, this benchmark model can outperform some of the density-based approaches.
\end{itemize}

\subsection*{Prediction performance metrics}
In order to compare the goodness of the predictions of the different approaches we use standard evaluation metrics, widely used in supervised machine learning \citep{hossin2015review}. As different metrics capture different aspects of the prediction problem, only the comparison between various measure of performance can provide a global view of the effectiveness of a forecast approach.\\
The elements that we want to predict are the possible activations, that is, those elements of $\textbf{M}^{2011}$ that were always zero in 2000-2009. The 0s are called negatives, and the 1s are called positives. The elements equal to 1 that are correctly predicted are called true positives (TP); and similarly one can define the false positives (FP), the true negatives (TN) and the false negatives (FN) as, respectively, the 0s predicted as 1s, the correctly predicted 0s, and the 1s predicted as 0s.\\
To evaluate the predictions done with the different approaches, we have used three evaluation metrics:
\begin{itemize}
    \item \textbf{Best-F1}: The F1 score is defined as the harmonic mean between precision and recall:
    \begin{equation*}
        \textrm{F1} = 2 \left(\frac{1}{\text{precision}(\tau)} + \frac{1}{\text{recall}(\tau)}\right)^{-1},
    \label{eq:F1}
    \end{equation*}
    where $\text{precision} = \frac{TP(\tau)}{TP(\tau)+FP(\tau)}$, and $\text{recall} = \frac{TP(\tau)}{TP(\tau)+FN(\tau)}$ are close to 1 of $FP$ and $FN$ are minimized, respectively. In such a way, the F1 score penalizes the errors in both sides.\\
    Note that in order to compute precision and recall one has to specify the scores? binarization threshold $\tau$, that is, the number above which the score is associated to a predicted 1. By adopting the Best-F1, we are considering the threshold parameter $\tau$ that maximizes, a posteriori, the F1 score. Note that the highest possible value of Best-F1 is $1$, which indicates that both precision and recall are equal to $1$, and the lowest possible value is $0$, if one of the precision and recall is zero.
    \item \textbf{Precision@100}: Here we focus on the top $100$ scores elements in $\textbf{S}^{2011}$: if the model is correct, many of these possible activations should become realized activations. The Precision@100 is the ratio between the number of how many of these 100 are true positives (that is, correctly predicted realized activations), and 100, i.e. the number of elements that we are considering. This represents a global assessment, that considers the score matrix as a whole.
    \item \textbf{mPrecision@10}: While the Precision@100 provides a global measure of the precision of the approach, we would like to have a measure of our average predictive performance for each firm. To do this, we evaluate the mPrecision@10. We consider the 10 best scores for each row, i.e. for each firm, and the compute the fraction of true positives. At the end, we average over the firms. Since most of the firms do not show at least 10 realized activations, the global number is far from 1. We have computed the mPrecision also restricting ourselves only to the firms with 10 or more realized activations, finding similar qualitative results.
\end{itemize}

\bibliography{Prediction_Comp_Tech__JMLR_}

\section*{Acknowledgements}
We thank Lorenzo Napolitano for providing the patent data used for this work and Andrea Tacchella for insightful discussions.



\end{document}